\documentclass[a4paper,12pt]{article}

\usepackage{amssymb,amsmath,empheq,epsfig,graphicx}

\usepackage{color}

\newcommand{\ag}{{\alpha_G}}             
\newcommand{\ampRid}{\mathfrak{M}}       
\newcommand{\atanh}{\mathrm{atanh}}      
\newcommand{\bk}[1]{\langle #1 \rangle}  

\newcommand{\bt}{{\boldsymbol{b}}}
\newcommand{\cut}{\mathrm{cut}}           
\newcommand{\dif}{\mathrm{d}}            
\newcommand{\el}{{\mathrm{el}}}          
\newcommand{\esp}[1]{\mathrm{e}^{#1}}    
\renewcommand{\hom}{\omega}               
\newcommand{\ic}{{\mathrm{ic}}}          
\newcommand{\match}{{\mathrm{matched}}}  
\newcommand{\Nor}{\mathcal{N}}           
\newcommand{\om}{\omega}
\newcommand{\omE}{\frac{\hom}{E}}
\newcommand{\tomE}{\textstyle{\frac{\hom}{E}}}
\newcommand{\ord}[1]{\mathcal{O}\left(#1\right)}
\newcommand{\Qt}{{\boldsymbol{Q}}}
\newcommand{\qt}{{\boldsymbol{q}}}
\renewcommand{\Pr}{\mathcal{P}}          
\newcommand{\po}{P}                      
\newcommand{\pp}{p}                      
\newcommand{\R}{\mathbb{R}}              

\newcommand{\regge}{\mathrm{Regge}}
\newcommand{\soft}{\mathrm{soft}}
\newcommand{\tb}{\bar{\tau}}             
\newcommand{\tfa}{{\cal M}}              
\newcommand{\Tht}{{\boldsymbol{\Theta}}}
\newcommand{\tht}{{\boldsymbol{\theta}}}
\newcommand{\ui}{\mathrm{i}}             
\newcommand{\vq}{\vec{q}}                
\newcommand{\xt}{{\boldsymbol{x}}}

\numberwithin{equation}{section}

\title{
{\bf Unitarity restoring graviton radiation\\ in the collapse regime of scattering}}

\author{
   Marcello~Ciafaloni
   \footnote{Email: ciafaloni@fi.infn.it}
   \\
   {\sl\small Dipartimento di Fisica, Universit\`a di Firenze and INFN Firenze}\\
   {\sl\small Via Sansone 1, 50019 Sesto Fiorentino, Italy}
   \\[1ex]
   and
   \\[1ex]
   Dimitri~Colferai
   \footnote{Email: colferai@fi.infn.it}
   \\
   {\sl\small Dipartimento di Fisica, Universit\`a di Firenze and INFN Firenze}\\
   {\sl\small Via Sansone 1, 50019 Sesto Fiorentino, Italy}\\[5mm]
   \\[5mm]
}
\date{}

\begin{document}

\maketitle

\begin{abstract}

We investigate graviton radiation in gravitational scattering at small impact
parameters $b<R\equiv 2G\sqrt{s}$ and extreme energies $s\gg M_P^2$, a regime in
which classical collapse is thought to occur, and thus radiation may be
suppressed also. Here however, by analyzing the soft-based representation of
radiation recently proposed in the semiclassical ACV framework, we argue that
gravitons can be efficiently produced in the untrapped region
$|\xt|\gtrsim R>b$, so as to suggest a possible completion of the unitarity
sum. In fact, such energy radiation at large distances turns out to
compensate and to gradually reduce to nothing the amount of energy $E'$ being
trapped at small-$b$'s, by thus avoiding the quantum tunneling suppression of
the elastic scattering and suggesting a unitary evolution. We finally look at
the coherent radiation sample so obtained and we find that, by energy
conservation, it develops an exponential frequency damping corresponding to a
``quasi-temperature'' of order $\hbar/R$, which is naturally related to a
Hawking radiation and is suggestive of a black-hole signal at quantum level.
\end{abstract}



\section{Introduction\label{s:intro}}

Gravitational scattering at extreme energies ($s\gg M_P^2$) and possibly small
impact parameters ($b\gtrsim R\equiv 2G\sqrt{s}$) was devised, from the
beginning~\cite{tHooft:1987rb,Muzinich:1987in,ACV87,GrMe87,ACV88,VeVe91,ACV90,ACV93},
as a probe of quantum gravity in a regime close to classical collapse.  More
recently, transplanckian scattering has been revived at both
classical~\cite{GrVe14} and quantum
level~\cite{Dvali:2014ila,CCV15,Addazi:2016ksu} with the purpose of describing
the radiation associated to extreme energies and of gaining in this way a better
understanding of the possibly collapsing system.

By analyzing string-gravity in the parameter region
$b\gg R\gg l_P\equiv\sqrt{G\hbar}$, ACV proposed a semiclassical
approach~\cite{ACV93} to gravitational scattering, based on an effective
action~\cite{Li91,VeVe91} which allows to compute the eikonal function
$\delta(\bt,s)$ as an all-order expansion in the parameter $R^2/b^2$. In the
axisymmetric reduced-action model they~\cite{ACV07} obtained
\begin{equation}\label{deltabs}
  \delta(\bt,s) = \ag \left[\log\frac{L}{R}+\Delta(\bt)\right] \;,\qquad
\Delta(\bt) \equiv -\atanh(t_b)+\frac12 - \frac1{2t_b} \;,
\end{equation}
where $t_b(b/R)$ is determined by the criticality equation
\begin{equation}\label{criticality}
  t_b(1-t_b^2) = \frac{R^2}{b^2}
\end{equation}
and $L$ is an IR cutoff, factorized in the elastic $S$-matrix expression
\begin{equation}\label{Sel}
  S_\el(\bt,s) = \esp{2\ui\delta}
  = \esp{2\ui\ag\log\frac{L}{R}}\esp{2\ui\ag\Delta(\bt)} \;.
\end{equation}
We notice that the leading contribution $\ag\log(R/b)$ is corrected in
$\Delta(\bt)$ by higher order terms, providing the ACV resummation, and that the
criticality equation~\eqref{criticality} identifies a branch-cut singularity of
the series at $b^2=b_c^2\equiv(3\sqrt{3}/2)R^2$, below which the eikonal
function acquires a (positive) imaginary part. As a consequence, starting from a
small impact parameter $b<b_c(R)$, the elastic channel acquires the suppression
factor
\begin{equation}\label{suppression}
  |S_\el(\bt,s)|^2 \simeq
  \begin{cases}
    \displaystyle\exp\left[-\frac{4\sqrt{2}}{3}\ag
      \left(1-\frac{b^2}{b_c^2}\right)^{3/2}\right]
    & \qquad(b_c-b\ll R) \\[6mm]
    \displaystyle\exp\left[-2\ag\left(\pi-\frac{3\sqrt{3}}{2}
        \left(\frac{b}{R}\right)^{2/3} \right)\right] &\qquad(b\ll R)
  \end{cases}
\end{equation}
which has the interpretation of tunneling probability~\cite{CC08} through a
repulsive Coulomb-type barrier in metric space, which is classically
forbidden. Eqs.~\eqref{deltabs}-\eqref{suppression} summarize the main
properties of the elastic ACV $S$-matrix~\eqref{Sel} for $b>b_c\sim R$ and
$b<b_c$, respectively.

More recently, the semiclassical (eikonal) framework has been revived in various
approaches at both classical~\cite{GrVe14} and quantum
level~\cite{CCV15,CCCV15,CC16} in order to describe the graviton radiation
associated to scattering. Since the typical graviton energy
$\om\sim 1/R \sim M_P^2/E \ll M_P$ is soft in the transplanckian regime
$(\ag\sim s/M_P^2\gg 1$), the radiation amplitude is well described by the
Weinberg current~\cite{We65} in the fragmentation region of phase space.
Furthermore, it has been noticed in~\cite{CCV15,CCCV15} that actually a unified
formulation of the amplitude applies to the central (Regge) region as well,
leading to the so-called ``soft-based representation'' of graviton emission.
Finally, the latter can be extended~\cite{CC16} to the ACV resummed formulation
of the eikonal mentioned before, in the ``reduced-action'' model.

Properties of the radiation sample generated in such a way in transplanckian
scattering have already been described in~\cite{CC16} by applying the soft-based
representation for both small and finite scattering angles ($b\geq b_c$). The
results emphasize the role of the gravitational radius $R$: by combining the
relatively small emitted energy ($\om/E \ll 1$) with the large-number
$n\sim\ag=ER$ of single-hits in eikonal scattering, the variable $\om R$ emerges
and identifies the main features of the energy emission distribution. The
outcome is an operator $S$-matrix combining the resummed-eikonal and radiation
in a (unitary) coherent state
\begin{align}
  \hat{S}&=\esp{2\ui\delta} \exp\left\{\int\frac{\dif^3 q}{\hbar^3\sqrt{2\omega}}\;
    2\ui\left[\sum_\lambda \ampRid_\lambda(\bt,\vq) a^\dagger_\lambda(\vq)
      +\text{h.c.}\right] \right\}\label{Shat} \\
  \frac{\ampRid_\lambda(\bt;\om,\qt)}{\esp{\ui\lambda\phi_\tht}} &=
  \sqrt{\ag}\frac{R}{\pi}\int\frac{\dif^2\xt}{2\pi|\xt|^2\esp{\ui\lambda\phi_\xt}}
  \frac{\esp{\ui\qt\cdot\xt}}{2\ui\om R}\left\{\esp{2\ui\om R\left[
        \Delta(\bt-\xt)-\Delta(\bt)\right]}-\esp{2\ui\om R\frac{E}{\om}\left[
        \Delta(\bt-\omE\xt)-\Delta(\bt)\right]} \right\} \label{ampRis}
\end{align}
with $\Delta$ given in eq.~\eqref{deltabs}, while $q$ and $\lambda$ denote
  graviton momentum and helicity, respectively.

We should remark, as a premise to the following investigation, that the unitary
form of eq.~\eqref{Shat} --- which is part of the proposal
in~\cite{CCV15,CCCV15,CC16} --- was based on appropriate and factorized
virtual corrections [eq.~\eqref{P0}], which are argued for in secs.~\ref{s:usbr}
and \ref{s:mge}, but are not really derived from some underlying quantum gravity
theory. For that reason we prefer to talk of ``unitarization procedure'' rather
than ``unitarity proof'' in passing from the tree-level amplitude $\ampRid$ to
the final unitary coherent state~\eqref{Shat}, with its factorized emission
structure.

Furthermore, we have shown~\cite{CC16} that such (approximate)
independent-particle picture can be extended to the approach-to-collapse regime
$b\to b_c^+$ of enhanced radiation by incorporating energy-conservation
constraints in the unitarization procedure. The latter cause the emergence of a
novel, exponential frequency damping whose coefficient $\tb R$ defines what we
call a ``quasi-temperature'' $T\equiv\hbar/(\tb R)$ which is naturally related
to the Hawking temperature~\cite{Hawking:1974sw,Hawking:2015qqa}. The main
difference, though --- and the motivation for its unconventional name --- is
that $\tb$ is supposed to keep quantum coherence, and is not due to a
statistical averaging.  Rather, the lack of sizeable correlations in such result
is due to the soft-graviton dynamics we started with, which led to the unified
form of the $b\to b_c^+$ emission amplitudes.

A basic question then arises: what about the collapse regime of $b<b_c$, in
which the energy $2E$ appears to be ``trapped'' because elastic unitarity is
exponentially violated by the suppression factors~\eqref{suppression} without
apparent contributions in the most naive radiation
models~\cite{CC08,CC09,CCF11,CC14}? In other words, is unsuppressed radiation
predictable for $b<b_c$ in our present soft-based representation? That is
precisely the question that we address in this paper. We shall argue that during
eikonal scattering, the soft-radiation process has indeed the ability to reduce
the amount of ``trapped'' energy crossing the barrier, and thus to gradually
eliminate (sec.~\ref{s:mge}) the suppression factor~\eqref{suppression}. As a
consequence (secs.~\ref{s:urqt},~\ref{s:icc}), our
coherent radiation sample can efficiently contribute to the unitarity sum, and
still may have a ``normal'' quasi-temperature of order $1/R$. For that reason,
it is a good candidate for the generalized unitarization procedure that we shall
describe, as discussed in sec.~\ref{s:d}.

\section{Unified soft-based representation of single-emission amplitude\label{s:usbr}}

We start, in the ACV framework, from the (irreducible) resummed eikonal
$\ag\Delta(\bt)$ in eq.~\eqref{deltabs}, whose Fourier transform defines a
``potential'' $\tilde\Delta(\Qt)$ in transverse space. In the soft limit and in
the fragmentation region, the emission amplitude is then given by the
external-line insertion formula, which factorizes in $\Qt$-space as follows
($\hbar=1$):
\begin{equation}\label{Msoft}
  \tfa^\soft_\lambda(\bt,E,\qt,\om) = \sqrt{\ag}\frac{R}{\pi}\int
  \frac{\dif^2\Qt}{2\pi}\;\tilde\Delta(\Qt)\esp{\ui\Qt\cdot\bt}
  \left[\frac{E}{\hom}\left(
      \esp{-\ui\lambda(\phi_{\qt-\frac{\hom}{E}\Qt} -\phi_\qt)}-1\right)\right]
  \;,
\end{equation}
where $\qt$ is the transverse momentum of the emitted graviton, $\phi_\qt$ is
its azimuth in the transverse plane, $\lambda = \pm2$ its helicity, and the
factor in square brackets comes from the explicit computation of the Weinberg
current on helicity states~\cite{CCV15}.

It was shown in~\cite{CCCV15} that a similar formula is able to describe graviton
emissions in the central region --- where the Lipatov current~\cite{Li91} should
be used in the Regge limit --- by just performing a simple subtraction of the
same expression at scale $E\to\hom$. The unified amplitude thus reads
\begin{align}
  \tfa_\lambda(\bt,E,\qt) &= \sqrt{\ag}\frac{R}{\pi}\esp{\ui\lambda\phi_\qt}
  \int\frac{\dif^2\xt}{2\pi|\xt|^2\esp{\ui\lambda\phi_\xt}}\;
  \esp{\ui\qt\cdot\xt} \nonumber \\
  &\quad\times \left[
    \frac{E}{\hom}\Big(\Delta\big(\bt-\tomE\xt\big)-\Delta(\bt)\Big) -
    \Big(\Delta(\bt-\xt)-\Delta(\bt)\Big)\right] \;, \label{uniamp}
\end{align}
where we have exchanged the $\Qt$ integration with an $\xt$ integration that
provides a convenient representation of phase-transfers (e.g.\ for $\lambda=-2$):
\begin{equation}\label{zRep}
  \esp{2\ui \phi_{\tht}} - \esp{2\ui \phi_{\tht'}}
  = -2 \int \frac{\dif^2\xt}{{2\pi x^*}^2} \left( \esp{\ui A \xt \cdot \tht}
    - \esp{\ui A \xt \cdot \tht'} \right) \;, \qquad(A\in \R^*) \;.
\end{equation}

We notice that eq.~\eqref{uniamp} is directly expressed in terms of the eikonal
function $\delta(\bt) \cong \ag\Delta(\bt)$ of eq.~\eqref{deltabs}, which occurs
in the modulating function
\begin{align}\label{PhiR}
  \Phi(\om,\xt) &\equiv \frac{E}{\hom}\left[\Delta(\bt-\tomE\xt)-\Delta(\bt)
    \right] -\left[\Delta(\bt-\xt)-\Delta(\bt)\right]
    \equiv \Phi_A(\xt) - \Phi_B(\xt) \;,
\end{align}
where the first (second) term is in correspondence with external (internal)
insertions. Thus, the single-exchange amplitude~\eqref{uniamp} measures the
Fourier transform of the ``soft-field''
\begin{equation}\label{hsoft}
  h_s^{(\lambda)}(\om,\xt) =
  -\frac{\Phi(\om,\xt)}{\pi^2|\xt|^2\esp{\ui\lambda\phi_\xt}} \;,
\end{equation}
which plays an important role in the expression of the ACV metric
also~\cite{CCCV15}.

\begin{figure}[ht]
  \centering
  \includegraphics[width=0.66\linewidth]{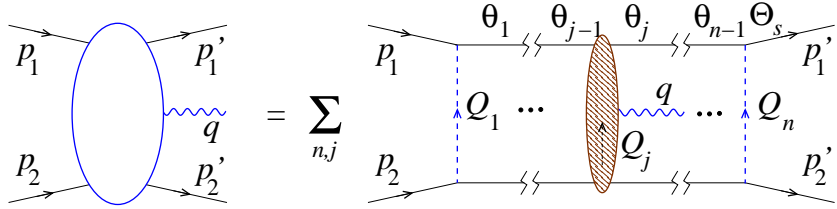}
  \caption{\it Graviton emission from the eikonal ladder. The $n$-rung diagram
    with the emission from the $j$-th exchange is denoted by $\tfa^{[n,j]}$ in
    the text.}
  \label{f:eikEmission}
\end{figure}

The full graviton radiation amplitude is now obtained by superimposing the
single-exchange amplitudes~\eqref{uniamp} over all the $n$ rungs of the eikonal
diagrams (fig.~\ref{f:eikEmission}). The contribution of the $j$-th rung can be
written as
\begin{align}
  \tfa^{[n,j]}_{\lambda}(\bt,E,\qt) &= \esp{\ui\lambda\phi_\tht}
  \sqrt{\alpha_G}\frac{R}{2} \frac{(2\ui\ag)^{n-1}}{n!} \times \nonumber \\
  &\quad\int\dif^2\xt \;\esp{\ui\qt\cdot\xt}
  \left[\Delta(\bt-\omE \xt)\right]^{j-1} h_s^{(\lambda)}(\om,\xt)
  \left[(E-\om)\Delta(\bt)+\om\Delta(\bt-\xt)\right]^{n-j} \;, \label{Mnj}
\end{align}
where we notice two important effects. Firstly, the $j$-th incidence angle is
rotated with respect to the $z$-axis, by translating the $\tht$-dependence by
the quantity $\Tht_j=(\Qt_1+\cdots+\Qt_{j-1})/E$. That produces in turn, after
Fourier transform, the shift $-(\om/E)\xt$ in the impact parameter of the
elastic amplitudes before emission.

Secondly, after emission we have the rescattering effects: the energetic
particle acquires the recoil energy $E-\om$, and the emitted graviton at
position $\xt$ rescatters with energy $\om$ and relative impact parameter
$\bt-\xt$.

Finally, the summation over $j$ of the contributions~\eqref{Mnj} is performed by
the formula
\begin{equation}\label{sum1}
  \sum_{n=1}^\infty \frac1{n!}\sum_{j=1}^n A^{j-1} B^{n-j}
  =\sum_{n=0}^\infty \frac1{n!}\frac{A^n- B^n}{A-B} 
  = \frac{\esp{A}-\esp{B}}{A-B} \;,
\end{equation}
where
\begin{equation}\label{phiAB}
  A \equiv 2\ui\ag\left[\Delta(\bt)+\omE\Phi_A(\xt)\right] \;, \qquad
  B \equiv 2\ui\ag\left[\Delta(\bt)+\omE\Phi_B(\xt)\right]
\end{equation}
are given in terms of the $\Phi_{A,B}$ of eq.~\eqref{PhiR}. We thus realize that
the factor $\Phi(\xt)=\Phi_A-\Phi_B$ in the soft field $h_s^{(\lambda)}$ cancels
out with the $A-B$ denominator of the summation~\eqref{sum1}. That cancellation
is conceptually surprising. Somehow, the identity
\begin{equation}\label{softRegge}
  \tfa_\match =
  \left.\soft\right|_E -\left.\soft\right|_{\hom}
  \simeq \left.\regge\right|_E
\end{equation}
--- that was interpreted as a decomposition of external plus internal insertions in
the soft language --- acquires now the interpretation of ``incidence-changing''
plus ``rescattering'' terms in the Regge language.

The final result can thus be written as
\begin{align}
  \tfa_{\lambda}(\bt,E,\qt) &= \esp{2\ui\ag\Delta(\bt)}
   \ampRid_{\lambda}(\bt,\om,\qt) \nonumber \\
  \ampRid_{\lambda}(\bt,\om,\qt)
  &= \sqrt{\alpha_G}\frac{R}{\pi}\esp{\ui\lambda\phi_\qt} \int
  \frac{\dif^2\xt}{2\pi|\xt|^2\esp{\ui\lambda\phi_\xt}} \;
  \esp{\ui \qt\cdot\xt} \esp{2\ui\om R\Phi_A(\xt)}
  \frac{\esp{-2\ui\om R \Phi(\xt)}-1}{2\ui\om R} \;.\label{ampRid2}
\end{align}

\section{Multi-graviton emission and unsuppressed radiation in the collapse
  regime\label{s:mge}}

\begin{figure}[ht]
  \centering
  \includegraphics[width=0.65\linewidth]{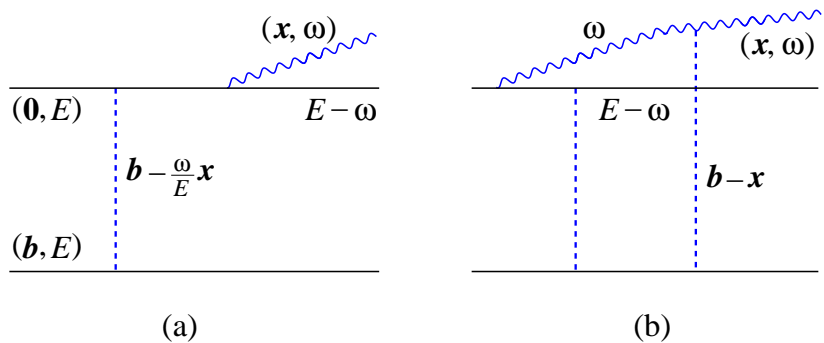}
  \caption{\it Radiation diagram of the soft-based representation: (a) External
    line insertion shifts the impact parameter $\bt-\omE\xt$ and rotates the
    incidence angle. (b) Internal line insertion scatters at recoil energy
    $E-\om$ and rescatters at position $\bt-\xt$ and energy $\om$.}
  \label{f:amplitude2to3}
\end{figure}

As we have just seen, the CC~\cite{CC16} method for incorporating the resummed
eikonal $\ag\Delta(\bt)$ of eq.~\eqref{deltabs} in the radiation process is best
illustrated (fig.~\ref{f:amplitude2to3}) by the one-graviton emission amplitude
\begin{equation}
  \tfa_{2\to3} = \int\frac{\dif^2\xt}{(2\pi)^2} \; \frac{\sqrt{\ag}}{x^{*2}}
  \frac{\esp{\ui \qt\cdot\xt}}{\ui\om} \left[ \esp{2\ui(E-\om)R\Delta(\bt)
      +2\ui\om R\Delta(\bt-\xt)} - \esp{2\ui\ag\Delta(\bt-\omE\xt)}
  \right] \label{M23}
\end{equation}
which contains {\it(i)} the rescattering term with its typical $(E-\om)$ recoil
energy (which is here listed as first term) and {\it(ii)} the incidence-changing
term with its $\om$-dependent shift (second term). From eq.~\eqref{M23}, by
dividing out the $2\to2$ $S$-matrix $\exp[2\ui\ag\Delta(\bt)]$, we obtain the
single-emission probability amplitude
\begin{align}
  \ampRid_\lambda(\bt,\om,\qt) \simeq \int\frac{\dif^2\xt}{(2\pi)^2} \;
  \frac{\sqrt{\ag}}{|x|^2\esp{\ui\lambda\phi_\xt}}
  \frac{\esp{\ui \qt\cdot\xt}}{\ui\om}
  &\Big\{ \esp{-2\ui\om R \left[\Delta(\bt)-\Delta(\bt-\xt)\right]} -1
       \nonumber\\
      & +1 - \esp{2\ui\ag\left[\Delta(\bt-\omE\xt)-\Delta(\bt)\right]}
        \Big\} \label{ampRid} \;,
\end{align}
where in curly brackets we have singled out the rescattering and
incidence-changing terms.

Multiple emission is then treated by arguing~\cite{CCCV15} that, in the soft
region $\om_i \ll E$, real emission factorization holds in the form
\begin{equation}\label{factorization}
  \tfa_{2\to2+N} \simeq \esp{2\ui\ag\Delta(\bt)} \prod_{i=1}^N
  \ampRid_{\lambda_i}(\bt,\om_i,\qt_i) \times
  \left[1+\ord{\frac{\om_j^2}{E^2}}\right] \;.
\end{equation}
In the case $b>b_c$, $\Delta(\bt)$ is real and the
factorization~\eqref{factorization} can be extended to virtual corrections by
addition of a factor $\sqrt{\Pr_0}$ --- the no-emission amplitude ---, where 
\begin{equation}\label{P0}
  \Pr_0 = \exp\left\{-2\int\frac{\dif^3 q}{\om}\;\sum_\lambda
    |\ampRid_\lambda|^2\right\}
\end{equation}
is the no-emission probability. Eq.~\eqref{factorization}, with the correction
factor~\eqref{P0} is then equivalent to using the unitary $S$-matrix
parametrization
\begin{align}
  S &= \esp{2\ui\delta}\exp\left\{\int\frac{\dif^3 q}{\sqrt{2\om}}\;
    2\ui \sum_\lambda \left[ \ampRid_\lambda a^\dagger_\lambda(q)
      + \ampRid_\lambda^* a_\lambda(q) \right] \right\} \label{Smatrix} \\
  \om \po(\om) &= 2\sum_\lambda \int\dif^2\qt\; |\ampRid_\lambda|^2 \label{oPo}
\end{align}
corresponding to the ``linear'' coherent-state operator~\eqref{Shat} and to
its emission density.

In~\cite{CC16} we noticed that the independent-particle picture just outlined is
not fully consistent when scattering angle and coupling enter the collapse
region and radiation should be corrected for energy-conservation effects. The
latter can be taken into account by introducing, besides $\sqrt{\Pr_0}$, an
amplitude renormalization factor $1/\sqrt{\Nor(E)}$, where $\Nor(E)$ is
dependent on the available energy $E$ and is determined by unitarity, by
including the kinematical constraints event by event. The outcome is then the
occurrence of the quasi-temperature factor $\esp{-\om/T}$ in inclusive
distributions~\cite{CC16}, which allows the correspondence of our coherent
radiation sample with a Hawking radiation.

The case $b<b_c$  is deeply different, however. In fact, in such case
$\Delta(\bt)$ acquires an imaginary part which is of size $\ui\pi/2$ in the
$b\ll R$ limit of deep collapse and has the interpretation of tunneling through
a barrier~\cite{CC08}. As a consequence, the elastic amplitude in the
Ansatz~\eqref{factorization} is exponentially suppressed like
$\esp{-\pi\ag}=\esp{-\pi ER}$ (for $E=\sqrt{s}/2$), and thus the combined use of
eq.~\eqref{factorization} and \eqref{P0} cannot possibly correspond to a unitary
parametrization of type~\eqref{Smatrix}. How to reach unitarity then?

Our purpose here is to single out those emission processes which, on the basis
of~\eqref{M23} and \eqref{ampRid} are not suppressed and can possibly lead
to unitarity recovery. We start noticing that, even at small impact parameters
$b\ll R$, the emission amplitude~\eqref{ampRid} stays unsuppressed
in~\eqref{factorization} if the graviton is emitted at sufficiently large
$|\xt|$, outside the trapped region, so that the corresponding eikonal is
real-valued. For the rescattering terms (with $\Delta(\bt-\xt)$) such "exit''
occurs already at $|\xt|>R\gg b$, while for the incidence-changing terms (with
$\Delta(\bt-\omE\xt)$) we have to require that $|\xt|>(E/\om)R$ be much
larger, in which case the $\xt$-integration is suppressed by a phase space
factor $\om^2/E^2$ in the interesting soft region $\om R\ll\ag$.

Therefore, for $b\ll R$, the rescattering terms appear to provide the best
visible radiation window and will be investigated firstly at multi-graviton
level in the following. Indeed, according to sec.~2 of~\cite{CC16}, the
multi-graviton factorization~\eqref{factorization} appears to hold exactly in
the rescattering case --- if incidence-changing terms are turned off --- because
the $\om_j R$-dependence is simply additive. Furthermore, the probability
amplitude $\ampRid$ of the rescattering term in~\eqref{ampRid} is increasing
like $\exp(\pi\om R)$
\footnote{Here we treat in detail, for definiteness, the $b\ll R$ case, but our
  arguments about suppression and compensation are valid for general $b<b_c$ by
  just replacing $\pi$ by $2\Im\Delta(b)$.}
within the kinematical bounds and thus yields a strong
radiation enhancement. The physical reason of such increase is just energy
conservation, because the recoil energy $(E-\om)$ signals the corresponding
suppression decrease.

As a consequence, by collecting all exponential terms, the independent
particles' distribution would read, approximately,
\begin{equation}\label{dPN}
  \dif\Pr(\{\om_j,N_j\}) = \Pr_0(E) \Theta(E-\sum_j\om_j N_j) \prod_{j}
  \frac{\left[\po(\om_j)\Delta\om_j\right]^{N_j}}{N_j!} \;,
\end{equation}
where
\begin{equation}\label{P0E}
  \Pr_0(E) = \esp{-2\pi ER} \;, \qquad
  \po(\om_j) \equiv \esp{2\pi\om_j R} \pp(\om_j)
\end{equation}
and $\pp(\om)\sim \ag/(\pi\om^3 R^2)\times[1-\ord{\esp{-2\pi\om R}}]$ is
obtained from the $\xt$-integration of the leading rescattering term with
important subleading corrections for $\om R\lesssim 1$. In more detail,
referring to one jet, we have from eq.~\eqref{ampRid} --- by use of the Parseval
identity --- the estimate of the rescattering density for
$|\xt|\gtrsim\ord{R}\gg|\bt|$
\footnote{Integrating over $\qt$ phase space in eq.~\eqref{oPo} limits
  $|\qt|=\om\sin\theta<\om=\ord{1/R}$ and thus introduces a natural cutoff
  $|\xt|>R$ in eq.~\eqref{omPom} for the validity of the Parseval identity at
  small $x$-values.}
\begin{align}
  \om\po(\om) &= \int\sum_\lambda|\ampRid_\lambda|^2 2\om^2\dif\Omega
  = \ag\int\frac{\dif^2\xt\;4(2\pi)^2}{(2\pi|\xt|)^4\om^2}
  \left| \esp{\om R[\pi+2\ui\Delta(-\xt)]}-1 \right|^2 \nonumber \\
  &\simeq \frac{\ag}{\pi\om^2 R^2}\esp{2\pi\om R}
  \left[(1-\esp{-\pi\om R})^2+4\bk{\sin^2\om R\Delta(\xt)}\esp{-\pi\om R}
  \right] \nonumber \\
  &\equiv \om\pp(\om)\esp{2\pi\om R} \;, \label{omPom}
\end{align}
where $\bk{\cdots}$ denotes $\xt$-integration (averaging).  We thus realize by
eq.~\eqref{dPN} that the suppression factor $\Pr_0(E)$ is compensated by the
rescattering enhancement factors $\esp{2\pi\om_j R}$ in $\po(\om_j)$ provided
$\sum_j \om_j N_j = E$, that is close to the energy conservation boundary in
which the whole energy $E$ is radiated off. That compensation may occur for a
few hard gravitons as well as a bunch of soft ones with $\om_j R=\ord{1}$, thus
allowing in principle a unitary behaviour with a normal quasi-temperature of
order $1/R$.

\section{Unitarity restoration and quasi-temperature:\\
  the rescattering terms\label{s:urqt}}

In order to better understand how suppression is avoided and unitarity is
possibly restored in the rescattering case, note that the
expressions~\eqref{dPN} and \eqref{omPom} allow the use of the energy-conserving
unitarization method based on the $\Nor(E)$ rescaling~\cite{CC16,GriVe_pc,AGK72}
in the $b<b_c$ case also.  In fact we can replace the independent-particle
distribution~\eqref{dPN} by
\begin{equation}\label{dPNt}
  \dif\tilde{\Pr}(\{\om_j,N_j\}) = \frac{\Pr_0(E)}{\Nor(E)}
  \int_{c-\ui\infty}^{c+\ui\infty} \frac{\dif\lambda}{2\pi\ui\lambda}\;
  \esp{\lambda(E-\sum_j \om_j N_j)}
  \prod_{j} \frac{\left[\po(\om_j)\Delta\om_j\right]^{N_j}}{N_j!} \;,
\end{equation}
where $\Pr_0/\Nor$ plays a role similar to the $b>b_c$ case, but
$\Pr_0=\esp{-2\pi ER}$ is the ACV-resummed suppression result.

The unitarity requirement $\sum_{\{N_j\}}\dif\tilde{\Pr}(\{\om_j,N_j\})=1$ (with
the kinematical constraint $\sum_j \om_j N_j \leq E$) determines $\Nor(E)$ as
\begin{align}
  \Nor(E) &\equiv \Pr_0(E) \left[ 1+\sum_{N=1}^\infty\frac1{N!} \int
  \left( \prod_{j=1}^N\dif\om_j P(\om_j) \right)
  \Theta(E-{\textstyle\sum_{j=1}^N\om_j}) \right] \nonumber\\
  &= \int_{2\pi R-\ui\infty}^{2\pi R+\ui\infty}
  \frac{\dif\lambda}{2\pi\ui\lambda}\; \Pr_0(E) \esp{\lambda E +\int_0^\infty
    \dif\om\;\esp{-\lambda\om} \po(\om)} \;, \label{NE}
\end{align}
where we have set $\Re\lambda\geq c = 2\pi R$ to let the exponent integrand to
formally converge. By the translation $\lambda = (2\pi+\tau)R$ we then obtain
\begin{equation}\label{NEtau}
  \Nor(E) = \int_{-\ui\infty}^{+\ui\infty}\frac{\dif\tau}{2\pi\ui(\tau+2\pi)}\;
  \esp{\tau RE+\int_0^\infty\dif\om\;\esp{-\tau\om R}\pp(\om)} \;,
\end{equation}
where $\Pr_0$ has been replaced by 1 and $\po(\om)$ by $\pp(\om)$ of
eq.~\eqref{omPom}.%
\footnote{
  In writing~\eqref{NE} (\eqref{NEtau}) we have to exchange the order of
  $\lambda$ ($\tau$)-integrations with $\om$-integrations. That is simply
  achieved by the truncation of $\po(\om)$ at the kinematical boundary:
  $\po(\om)\to\po(\om)\Theta(\sqrt{s}/2-\om)$. That truncation is understood in
  the following, and is a consequence of the kinematical constraints also for
  $E\leq\sqrt{s}/2$.}
Therefore, all suppression and enhancement factors are now
eliminated and the determination of the energy-conserving $\dif\tilde{\Pr}$'s
can proceed as for $b>b_c$. Firstly, the distribution~\eqref{dPNt} in the
$\tau$-representation is
\begin{equation}\label{dPNtau}
  \dif\tilde\Pr = \frac1{\Nor(E)}\int_{\epsilon-\ui\infty}^{\epsilon+\ui\infty}
  \frac{\dif\tau}{2\pi\ui(\tau+2\pi)} \esp{\tau R (E-\sum_j \om_j N_j)}
  \prod_j \frac{[\pp(\om_j)\Delta\om_j]^{N_j}}{N_j !} \;,
\end{equation}
where $\Nor(E)$ in~\eqref{NEtau} can be estimated by a saddle point method.
Actually, in the simple example $\pp(\om)= \ag R \hat\pp = \text{const}$, we
find directly $\Nor(E)\sim\exp[2\sqrt{\ag E R \hat\pp}]$ showing that
$\Nor(E)\gg1$ despite the $\Pr_0(E)$ factor in~\eqref{NE}. That suggests the
direct interpretation of $\Nor(E)$ as a sort of entropy of the ``trapped
energy'' fragmentation process.

Secondly, the general saddle point equation reads
\begin{align}
  \ag &= \frac1{\tb+2\pi} + \ag F(\tb) \label{sadpoi} \\
  F(\tb) &\equiv \int_0^\infty \dif\om\;\frac{\om}{E}\esp{-\tau\om R} \pp(\om)
  \nonumber \\
  &= \int_0^\infty\dif(\om R)\;\frac{\ag}{\pi(\om R)^2}
  \left[(1-\esp{-\pi\om R})^2+4\bk{\sin^2\om R\Delta(\xt)}\esp{-\pi\om R}
  \right] \esp{-\tb\om R} \;, \nonumber
\end{align}
which for $\ag\gg 1$ leads to approximately $F(\tb)=1$, that is to
$\sum_j \om_j N_j = E$, or radiation of the total ``trapped'' energy.
Furthermore, we can calculate from eq.~\eqref{dPNt} the single-graviton
inclusive distribution $\dif N/\dif\om$, which involves fixing $\om_j=\om$ for
some $j$, and integrating over the remaining ones at $E-\om$ fixed.  It is
straightforward to see that this provides --- apart from small fluctuation
corrections $\ord{1/\ag}$ ---
\begin{equation}\label{dNdom}
  \frac{\dif N}{\dif\om} = \pp(\om)\frac{\Nor(E-\om)}{\Nor(E)}
  \simeq \pp(\om) \esp{-\tb R\, \om} \;,
\end{equation}
which is consistent with the average emitted energy in~\eqref{sadpoi} and,
together with eqs.~\eqref{NEtau}-\eqref{sadpoi}, appears to solve the $b\ll R$
model with a meaningful unitarization.

The exponential damping at large $\om R$, which affects all the inclusive
distributions as in eq.~\eqref{dNdom}, allows us to interpret $(\tb R)^{-1}$ as
the quasi-temperature of graviton radiation, and provides an effective cutoff in
frequency, superimposed to the density $\pp(\om)$, typical of our soft-based
representation. The overall picture of the ``trapped-energy'' fragmenting into
soft gravitons according to the distribution~\eqref{dPNtau} looks generically
compatible with ideas discussed in refs.~\cite{Dvali:2014ila} and
\cite{Addazi:2016ksu}, although applied --- in our case --- to the precise ACV
framework at fixed impact parameter $b$ and with $s$-channel iteration.  Arguments
for a cutoff are given also in the approach of ref.~\cite{Addazi:2016ksu} to the
transplanckian scattering without impact parameter identification of
ref.~\cite{Dvali:2014ila}.

\begin{figure}[ht]
  \centering
  \includegraphics[width=0.4\linewidth,angle=-90]{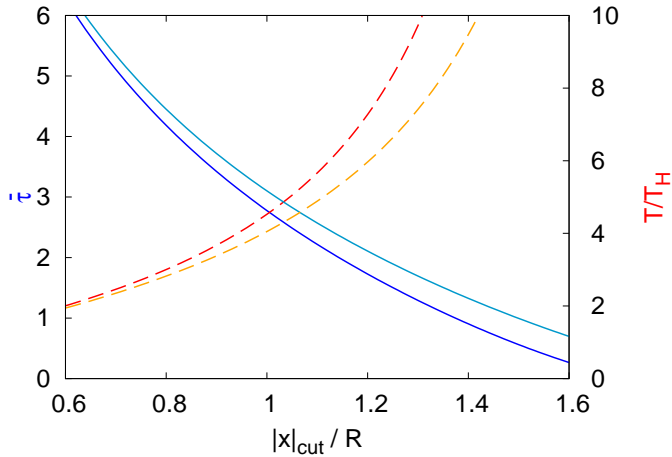}
  \caption{\it Dependence of the saddle-point value $\tb$ (solid-blue, left
    axis) and of the rescattering quasi-temperature $T\equiv 1/(\tb R)$
    (dashed-red, right axis) in units of $T_H\equiv 1/(4\pi R)$ on the cutoff
    parameter $|\xt|_\cut$ introduced in the approximate determination of the
    saddle-point eq.~\eqref{sadpoi}. Including incidence-changing effects at
    first order (see sec.~\ref{s:icc}) the saddle-point values increase
    (solid-light-blue) and correspondingly the quasi-temperature decreases
    (dashed-orange).}
  \label{f:taubarVSxcut}
\end{figure}

The precise determination of the inverse quasi-temperature $\tb$ for
rescattering requires a small-$\xt$ cutoff parameter $|\xt|_\cut = \ord{R}$
which takes contributions from $\Im\Delta(\xt) > 0$ if $|\xt|_\cut^2 < b_c^2 =
(3\sqrt{3}/2) R^2$, region where the $\xt$-dependence starts being suppressed
in~\eqref{sadpoi}. We then find the numerical results of
fig.~\ref{f:taubarVSxcut}, showing that $\tb \simeq 3$ is a reasonable estimate
(solid-blue curve), to be compared with $\tb=1.2$ from $b\to b_c^+$~\cite{CC16}.
We should notice, however, that $\tb$ is rather sensitive to the value of
$|\xt|_\cut$ around and below $b_c$. That means that the inclusion of
incidence-changing contributions to radiation may possibly be needed to provide
a more stable temperature estimate and a firmer conclusion on the unitary
behaviour on the basis of eqs.~\eqref{NEtau} and \eqref{sadpoi}.

We should add finally that, starting from eq.~\eqref{M23}, we can also compute
the probability distribution for the residual energy of the ``trapped''
gravitons $E'=E-\sum_j \om_j N_j$. By introducing the $\delta$-function
constraint, we get the formula
\begin{align}
  \frac{\dif\Pr}{\dif E'} &= \frac{\Pr_0(E)}{\Nor(E)} \sum_{\{N_j\}}
  \delta(E-E'-\sum_j \om_j N_j) \prod_j \frac{[\po(\om_j)\Delta\om_j]^{N_j}}{N_j!}
  \nonumber \\
  &= \frac{\Theta(E')}{\Nor(E)}\esp{-2\pi R E'} \int_{-\ui\infty}^{+\ui\infty}
  \frac{\dif\tau' R}{2\pi\ui}\; \esp{\tau'(E-E')R +\int_0^\infty \dif\om\;
    \pp(\om)\esp{-\om R\tau'}} \;, \label{dPdEp}
\end{align}
which can be estimated by a saddle point method also.

We note that the unitarity condition
$\int_0^\infty \dif E'\;\dif\Pr/\dif E'=1$ fixes again $\Nor(E)$ as in
eq.~\eqref{NEtau} by reproducing the translation $\tau'\to\tau'+2\pi$ in the
denominator. Indeed, the $E'$ distribution is mostly dependent on the tunneling
exponent $2\pi R$, which appeared in $\Pr_0(E)$. At $E'$ fixed the saddle point
$\tb'$ is determined by
\begin{equation}\label{tbp}
  F(\tb') = 1-\frac{E'}{E} \quad \left( \simeq 1 - \frac1{\ag(\tb+2\pi)} \right)
\end{equation}
and the average $\bk{E'}\simeq 1/(2\pi R+\tb R)$ makes $\tb'$ roughly consistent
with $\tb$ in eq.~\eqref{sadpoi}, with a maximal residual ``bound'' energy of
order $1/(2\pi R)$. It seems therefore that both the ``bulk'' temperature
$1/(\tb R)$ and the residual-energy temperature $1/(2\pi R)$ (reminiscent of the
Schwarzschild black-hole limit~\cite{ACV07}) play a role in this model. We
stress the point, however that, because of~\eqref{dPdEp}, they are consistent
with each other, due to their different definition.

\section{Incidence-changing contributions\label{s:icc}}

We have so far considered the rescattering terms in~\eqref{M23} and
\eqref{ampRid}, because they offer the first ``exit'' window at
$|\xt|\gtrsim R$. Now we want to estimate the incidence-changing window at
varying values of $\omE|\xt|$, in which the parameter $\om/E$ affects directly
the $\xt$-dependence.

The single-emission density of eq.~\eqref{omPom} becomes, approximately,
\begin{equation}\label{omPapp}
  \om\po(\om) \simeq \frac{\ag}{\pi\om^2\bk{\xt^2}} \left|
    \esp{\om R[\pi+2\ui\Delta(-\xt)]} -
    \esp{\ag[\pi+2\ui\Delta(-\omE\xt)]} \right|^2 \;,
\end{equation}
while the no-emission probability is $\Pr_0=\esp{-2\pi\ag}$ as usual for
$b\ll R$. We note that in the incidence-changing term the enhancement occurs at
coupling $\ag > \om R$, but there is further suppression, due to the $\om/E$
dependence in the eikonal expansions at small $\om|\xt|$ values
($\frac{b}{R} \ll \frac{\om|\xt|}{ER} \ll 1$) \cite{ACV07}
\begin{align}
  2\Delta\big(-\frac{\om}{E}\xt\big) &\simeq \ui\pi - 3 \esp{\ui\pi/3}
  \left(\frac{\om|\xt|}{ER}\right)^{2/3} +
  \ord{\frac{b^2}{R^2}}^{2/3} \;, \label{delomex} \\
  \esp{\ag\big[\pi+2\ui\Delta\big(-\omE\xt\big)\big]} &\simeq
    \esp{a\ag\left(\frac{\om R}{\ag}\right)^{2/3}} \;, \qquad
    \Big(a = \frac{3\sqrt{3}}{2}~\text{for}~\bk{\xt^2} = R^2\Big) \;.
    \label{defa}
\end{align}

We thus realize, by~\eqref{omPapp} and~\eqref{defa}, that we should consider two
regions for incidence-changing terms. In the region $|\xt|\geq \omE R$ the
trapping suppression is indeed canceled out, but the outcome is reduced by the
phase-space factor $(\om/E)^2$, yielding small ($\sim 1/\ag$) contributions in the
hard-graviton corner. On the other hand, in the region $|\xt|\gtrsim R$ we are
interested in, there is a small-$\om R$ configuration
$(\om R/\ag)^{1/3} < a/\pi < 1$ in which the enhancement~\eqref{defa} --- though
insufficient by itself to overcome the damping --- is anyway larger than the
single-density rescattering contribution, and therefore should be taken into
account. To this purpose, we can provide from~\eqref{defa} a crude estimate of
the Laplace transform
\begin{equation}
  \tilde{\po}_\ic(\lambda) \equiv \int_0^\ag \dif\om\;
  \po_\ic(\om)\esp{-\om\lambda} \simeq
  \begin{cases}
    \esp{\frac23 a\ag\left(\frac{4a}{3\lambda}\right)^2} &\qquad
    \left(\frac{4a}{3\lambda} < 1\right)\\
    \esp{(2a-\lambda)\ag} & \qquad \left(\frac{4a}{3\lambda} \geq 1\right)
  \end{cases}
  \;, \label{ptilde}
\end{equation}
where the last expression is provided by a saddle-point at
$\bar\om R=\ag\left(\frac{4a}{3\lambda}\right)^3$, if below $\bar\om R=\ag$, and
by the end-point $\om R=\ag$ otherwise. We note that
\begin{equation}\label{ppic}
  -\frac{\tilde\po'_\ic(\lambda)}{\tilde\po_\ic(\lambda)} =
  \begin{cases}
    \ag\left(\frac{4a}{3\lambda}\right)^3 = \bar\om R & \qquad
    \left(\frac{4a}{3\lambda} < 1\right) \\
    \ag & \qquad \left(\frac{4a}{3\lambda} \geq 1\right)
  \end{cases} \;.
\end{equation}

Our task, however, should be to provide a reliable estimate of the
incidence-changing terms at the many-graviton level required by
eq.~\eqref{dPNtau}, and that raises a variety of questions, involving both
matter of concept (factorization is justified for rescattering terms only) and
technical approximations for $b\ll R$.  Therefore, further analysis is needed
and treating incidence-changing contributions in detail in addition to
rescattering ones is outside the scope of the present paper.

We only point out that the rescattering estimate of $\Nor(E)$ in the
$\tau$-representation~\eqref{NEtau} can be improved by including
incidence-changing-effects at the level of single power of $\po_\ic(\om)$. In
that case the exponent in the integrand of~\eqref{NEtau} is corrected by adding
a $\log\tilde\pp_\ic(2\pi+\tau)$ from~\eqref{ptilde} and the saddle point
equation~\eqref{sadpoi} becomes
\begin{equation}\label{icsp}
  \ag = \frac1{\tb+2\pi} + \ag \left[ F(\tb) + \left(\frac{4a}{3(\tb+2\pi)}
      \right)^3 \right] \;,
\end{equation}
where the logarithmic derivative is taken from~\eqref{ppic}. The meaning
of~\eqref{icsp} is that a fraction of the overall energy is now radiated by
$\po_\ic$ also. The addition of such incidence-changing effect on the
saddle-point and temperature values is shown in fig.~\ref{f:taubarVSxcut}. There
is a moderate increase in $\tb$ (solid-light-blue curve) and a corresponding
decrease in temperature (dashed-orange curve).

\section{Discussion\label{s:d}}

Here we have investigated the collapse regime of gravitational scattering at
extreme energies ($E\gg M_p$) and small impact parameters ($b<b_c\sim R$), and
we have pointed out that --- in the present approach --- graviton radiation is
not necessarily suppressed, and actually multi-graviton amplitudes suggest how a
unitary $S$-matrix may still be found.

Our framework is semiclassical scattering~\cite{ACV93} in the
ACV-resummed~\cite{ACV07} eikonal formulation and in the soft-based
representation~\cite{CC16} of graviton radiation. Our suggestion is based on two
main points. Firstly, starting from jet energy $E=\sqrt{s}/2$ at impact
parameter $b\ll R\equiv 4GE$ elastic scattering to free-particle states is
exponentially suppressed by $|S_\el(b)|\simeq \esp{-\pi ER}$ because of
$\Im\Delta(b) \simeq \pi/2$ [eq.~\eqref{suppression}]. But rescattering
contributions of an emitted graviton at position $\xt$ is regulated by
$\Delta(\bt-\xt)$ [eq.~\eqref{M23}] and is thus unsuppressed if
$|\xt|^2>b_c^2\equiv(3\sqrt{3}/2)R^2\gg b^2$. That large-distance radiation is
in turn associated to the recoil energy $E-\om$ of the energetic particle, so
that the suppression amplitude is reduced --- and the emission one is enhanced
--- by the factor $\esp{\pi\om R}$, within the kinematical bounds. The same
conclusion is reached for general $b<b_c$ by replacing $\pi$ by
$2\Im\Delta(b)$.

Secondly, that suppression-enhancement correspondence goes through to
multi-graviton states provided the related soft-graviton ($\om_j\ll E$)
amplitudes are factorized, as argued for in~\cite{CC16} on the basis of eikonal
factorization, at least for rescattering amplitudes. In such a case, the overall
suppression factor in multi-graviton emission becomes
\begin{equation}\label{mulgrasup}
  \exp\Big[-\pi\big(E-\sum_{j=1}^N\om_j\big)R\Big]
  \Theta\big(E-\sum_{j=1}^N\om_j\big)
\end{equation}
and is therefore $\ord{1}$ (meaning no suppression) if $E=\sum_j\om_j$ or, in
other words, if the whole energy is radiated off. That may happen for a few hard
gravitons, but also for a bunch of soft ones, thus allowing in principle
unsuppressed emission amplitudes with a normal quasi-temperature of order $1/R$.

Therefore, if we take multi-graviton factorization for granted, the above
argument hints at the probability distribution~\eqref{dPNtau}, which is
unitarized by the normalization factor $\Nor(E)$ in~\eqref{NEtau} and is
characterized by the saddle point $\tb$ in~\eqref{sadpoi}, the inclusive
distribution~\eqref{dNdom} and thus the quasi-temperature $1/(\tb R)$.

That looks as the right path to follow in general, but unfortunately we do not
quite understand how to combine the incidence-changing contributions with the
rescattering ones at the multi-graviton level required by~\eqref{mulgrasup},
because of their uncertain factorization properties. That is not surprising,
because the incidence-changing terms (which regulate the rotation of the
incidence axis) are basically dependent on the overall coupling $\ag$ of the
energetic particles, while their $\om_j R$-dependence is tied up with the
$\xt$-dependence and is normally non-factorizable.

For those reasons, in sec.~\ref{s:urqt} we concentrate on the rescattering terms
(for which multi-graviton factorization is justified) by introducing a cutoff
$|\xt|_\cut=\ord{R}$, to regulate small-$\xt$ contributions. We thus find that
the unitarization method suggested by~\eqref{mulgrasup} actually works, with
some cutoff dependence of $\tb$ with $\tb\simeq 3$ for $|\xt|_\cut = R$. We feel
that a better understanding of incidence-changing contributions could be able to
close the gaps and reduce the cutoff dependence, as suggested by the provisional
estimate of fig.~\ref{f:taubarVSxcut}. Such analysis is deferred to further
investigations.

A perhaps more fundamental question to be discussed is what our results in
sec.~\ref{s:urqt} actually mean for the gravitational scattering and, possibly,
for black-hole physics. We have already noticed that there are two ways the
``trapped energy'' $\sqrt{s}=2E$ can be observed: either in full, without
accompanying soft-gravitons, by the amplitude $\sim\esp{-\pi\ag}$ (corresponding
to the residual energy temperature $1/(2\pi R)$) which is exponentially
suppressed, or instead by a sort of collective fragmentation into
soft-gravitons, described by the distribution~\eqref{dPNtau} and the
quasi-temperature $1/(\tb R)$. Shall we say that the latter is the most probable
issue and that, therefore, the unitary distribution~\eqref{dPNtau} represents
the quantum black-hole spectrum?

If that is really the case, then the solution of the unitarity problem and,
perhaps, of the information paradox would rely only on our ability to keep track
of the phases and to describe the quantum states. That brings us back to the
previous question of whether or not we are able to disentangle the full
multi-graviton amplitudes, and that, finally, seems to be matter of technique
and not matter of principle.

To conclude, we are aware of the fact that our framework is not a consistent
quantum gravity theory and is thus providing a limited description of
gravitational processes. Nevertheless, our discussion suggests that some
difficulties previously found with unitarity and the information paradox may be
solved by our proposal in a simpler way than previously thought.

\section{Acknowledgments}

It is a pleasure to thank Gabriele Veneziano for a number of interesting
conversations on the topics presented in this paper. We also wish to thank the
{\em Galileo Galilei Institute for Theoretical Physics} for hospitality while
part of this work was being done.

\bibliographystyle{h-physrev5}
\bibliography{collapse}

\end{document}